\documentclass[conference]{IEEEtran}

\usepackage{cite}
\usepackage{amsmath,amssymb,amsfonts}
\usepackage{amsthm}
\usepackage{algorithm, algpseudocode}
\usepackage{graphicx}
\usepackage{textcomp}
\usepackage{xcolor}
\usepackage{subcaption}
\usepackage{multirow}
\usepackage{booktabs}

\def\BibTeX{{\rm B\kern-.05em{\sc i\kern-.025em b}\kern-.08em
    T\kern-.1667em\lower.7ex\hbox{E}\kern-.125emX}}

\begin{document}

\title{Discovering Exfiltration Paths Using Reinforcement Learning with Attack Graphs}

\author{Tyler Cody$^{a}$$^{*}$,
        Abdul Rahman$^{b}$,
        Christopher Redino$^{b}$, \\
        Lanxiao Huang$^{a}$,
        Ryan Clark$^{b}$,
        Akshay Kakkar$^{b}$,\\
        Deepak Kushwaha$^{b}$,
        Paul Park$^{c}$, 
        Peter Beling$^{a}$,
        Edward Bowen$^{b}$\\
        \small $^{a}$National Security Institute, Virginia Tech \\
        \small $^{b}$Deloitte \& Touche LLP \\
        \small $^{c}$Deloitte Consulting LLP \\
        \small $^{*}$Corresponding Author: tcody@vt.edu \\
}

\IEEEoverridecommandlockouts
\IEEEpubid{\makebox[\columnwidth]{978-1-6654-2141-6/22/\$31.00~
\copyright2022
IEEE \hfill} \hspace{\columnsep}\makebox[\columnwidth]{ }}

\maketitle

\begin{abstract}

Reinforcement learning (RL), in conjunction with attack graphs and cyber terrain, are used to develop reward and state associated with determination of optimal paths for exfiltration of data in enterprise networks. This work builds on previous crown jewels (CJ)  identification that focused on the target goal of computing optimal paths that adversaries may traverse toward compromising CJs or hosts within their proximity.  This work inverts the previous CJ approach based on the assumption that data has been stolen and now must be quietly exfiltrated from the network. RL is utilized to support the development of a reward function based on the identification of those paths where adversaries desire reduced detection. Results demonstrate promising performance for a sizable network environment.

\end{abstract}


\section{Introduction}

The National Institute of Standards and Technology (NIST) special publication 800-53 revision 5 states that exfiltration\footnote{NIST 800-53r5 \cite{NIST800-53} states specifically that exfiltration lies within security control SC-07(10) for boundary protection to prevent unauthorized data movement (exfiltration).} (also called exfil) is the unauthorized movement of data within a network \cite{NIST800-53}. Many times, cyber attacks are considered successful if they exfiltrate data for monetary, disruptive, or competitive gain. Detection of exfiltration can be plagued with technical challenges as adversaries routinely encapsulate data within typically allowable protocols (e.g., http(s), DNS) which make it significantly harder to defend. Additionally, adversaries have been known to prefer traversing certain network paths for data theft to reduce detection and tripping cyber defenses so they do not raise suspicions. 

Heisting data requires two different plans: a plan to get to the data and a plan to exfiltrate the data without getting caught. Much effort in the cybersecurity industry is devoted to identifying and preventing points of weakness that allow authorized (i.e., adversarial) entry into a network. The most common exfiltration opportunity is moving data from a local network to an adversary network via the internet. To perform this, an adversary must gain access to the data on an organization's network, then send the data to a place off their network. Most organizations are focused on preventing network access, which leaves gaps in defenses for access from the network to the internet.

Much of the literature on automating penetration testing using RL has a focus on the way networks can be accessed (i.e., infiltration \cite{ghanem2018reinforcement, schwartz2019autonomous, ghanem2020reinforcement, chaudhary2020automated, yousefi2018reinforcement, chowdhary2020autonomous, hu2020automated, gangupantulu2021using, gangupantulu2021crown}). And while some consider using RL to detect exfiltration \cite{venkatesan2017detecting, albanese2018defending}, RL for conducting post-exploitation activities like exfiltration are under-studied \cite{maeda2021automating}. Maeda and Mimura apply deep RL to do exfiltration, however, they do not use a standard attack graph construct, but rather define states using an ontological model of the agent and define actions using task automation tools. \cite{maeda2021automating}. Their approach has several limitations:
\begin{itemize}
    \item The RL agent's inputs and outputs are greatly abstracted away from network structure, path structure, and cyber terrain, thereby limiting the ability to anchor agents to the \emph{real} computer network.
    \item The exfiltration methodology does not leverage automated frameworks for attack graph construction like MulVal \cite{ou2005mulval} or the vulnerability- and bug-reporting communities (e.g., via the Common Vulnerability Scoring System (CVSS) \cite{mell2007complete}).
    \item The output of the RL-based exfiltration method is not easily interpretable in terms of networks, their paths and configurations, and risks preferences regarding their traversal.
\end{itemize}
Whereas Maeda and Mimura's use of task automation tools  and ontologies make their proposed exfiltration method a highly automated means of actually performing exfiltration, this paper presents an alternative approach more tailored to automating exfiltration path discovery for cyber operator workflows.

This paper presents an RL method for discovering exfiltration paths in attack graphs. This paper proposes and combines:
\begin{enumerate}
    \item An approach for modeling service-based defensive cyber terrain in dynamic models of attack graphs.
    \item An RL-based algorithm for discovering the top-$N$ exfiltration paths in an attack graph.
\end{enumerate}
The presented methodology is aligned with a focus on network structure and configuration, path analysis, and cyber terrain. It maintains MulVal's focus on scalability and leverages the vulnerability- and bug-reporting communities via CVSS. Its outcomes can be directly understood as paths through networks, as is highlighted in a detailed discussion of the results. To support reproducibility, the RL solution methods, experimental design, and network model are specified in great detail. 

This paper is structured as follows: First, background on RL for penetration testing and on constructing Markov decision processes (MDPs)from attack graphs is given. After, the methods for modeling defensive terrain and discovering exfiltration paths are presented. Then, experimental design is described, results are presented, and findings are discussed. The paper concludes with remarks on modeling decisions, a synopsis, and a statement on future work.

\section{RL and Penetration Testing}

\subsection{Reinforcement Learning Preliminaries}

RL describes the paradigm of learning by interaction with an environment \cite{sutton2018reinforcement}. This contrasts directly with supervised learning where an oracle is queried for ground-truth labels. More formally, it describes a set of solution methods for approximate dynamic programming \cite{powell2007approximate}. It also addresses challenges associated with large and complex environments by approximating various aspects of planning and decision-making.

With respect to RL, agents learn by taking actions in environments $\mathcal{E}$ and receiving rewards. Commonly, environments $\mathcal{E}$ are modeled as MDPs. Finite MDPs are tuples $\{S, A, \Phi, P, R\}$ where $S$ and $A$ are states and actions, $\Phi \subset S \times A$ are admissible state-action pairs, $P:\Phi \times S \to [0, 1]$ is the probability transition function, and $R: \Phi \to \mathbb{R}$ where $\mathbb{R}$ are the reals is the expected reward function. An agent interacts with an environment $\mathcal{E} = \{S, A, \Phi, P, R\}$ by taking actions $a_t$ and receiving states $s_{t+1}$ and rewards $r_{t+1}$.

The learning procedure can be described in general terms as follows. Let $R_t$ be the discounted sum of future rewards,
\begin{equation}
    R_t = \sum_{k=0}^\infty \gamma^k r_{t+k},
\end{equation}
where $\gamma \in (0, 1)$ is a discount factor. The action value function $Q^\pi(s, a)$ can then be defined as
\begin{equation}
    Q^\pi(s, a) = \mathbb{E}[R_t|s_t=s, a],
\end{equation}
where $\pi$ is a policy mapping states and actions $(s, a)$ to the probability of picking action $a$ in state $s$.
The learning procedure aims to find the optimal action value function $Q^*(s, a)$,
\begin{equation}
    Q^*(s, a) = \max_\pi Q^\pi (s, a).
\end{equation}
Deep Q-learning (DQN) approximates $Q^*$ with a neural network $Q(s, a; \theta)$, where $\theta$ are parameters of the neural network \cite{mnih2013playing, mnih2015human}.

Alternatively, instead of learning the Q function, policies can be parameterized and learned directly. In policy gradient methods, the reward function is defined as
\begin{equation}
    J(\theta)=\sum_{s}d^\pi (s)V^\pi (s)=\sum_{s}d^\pi (s)\sum_{a}\pi_{\theta}(a|s)Q^\pi (s,a),
\end{equation}
where $d^\pi (s)$ denotes the stationary distribution of Markov chain for $\pi_{\theta}$. According to policy gradient theorem, the gradient $\nabla_{\theta}J(\theta)$ is given by
\begin{equation}
    \nabla_{\theta}J(\theta)\propto\sum_{s}d^\pi (s)\sum_{a}Q^\pi (s,a)\nabla_{\theta}\pi_{\theta}(a|s).
\end{equation}
The policy gradient theorem provides a basis for learning a parameterized policy. However, it suffers from high variance of gradient and instability. To overcome this, the value of the state $V^\pi (s)$, the value of using policy $\pi$ in state $s$ is introduced as the baseline:
\begin{equation}
    A^\pi (s,a)=Q^\pi (s,a)-V^\pi (s)
\end{equation}
where $A^\pi (s,a)$ is called the advantage. And the gradient is now given as
\begin{equation}
    \nabla_{\theta}J(\theta)=\mathbb{E}_{s\sim\rho^\pi,a\sim\pi_{\theta}}[\nabla_{\theta}\log\pi_{\theta}(s,a)A_{\pi}(s,a)]
\end{equation}
These gradients serve as the basis for the advantage actor-critic method (A2C), a standard policy gradient method in deep reinforcement learning \cite{mnih2016asynchronous}.

\subsection{RL for Penetration Testing}
\begin{figure}[t]
    \centering
    \includegraphics[width=0.45\textwidth]{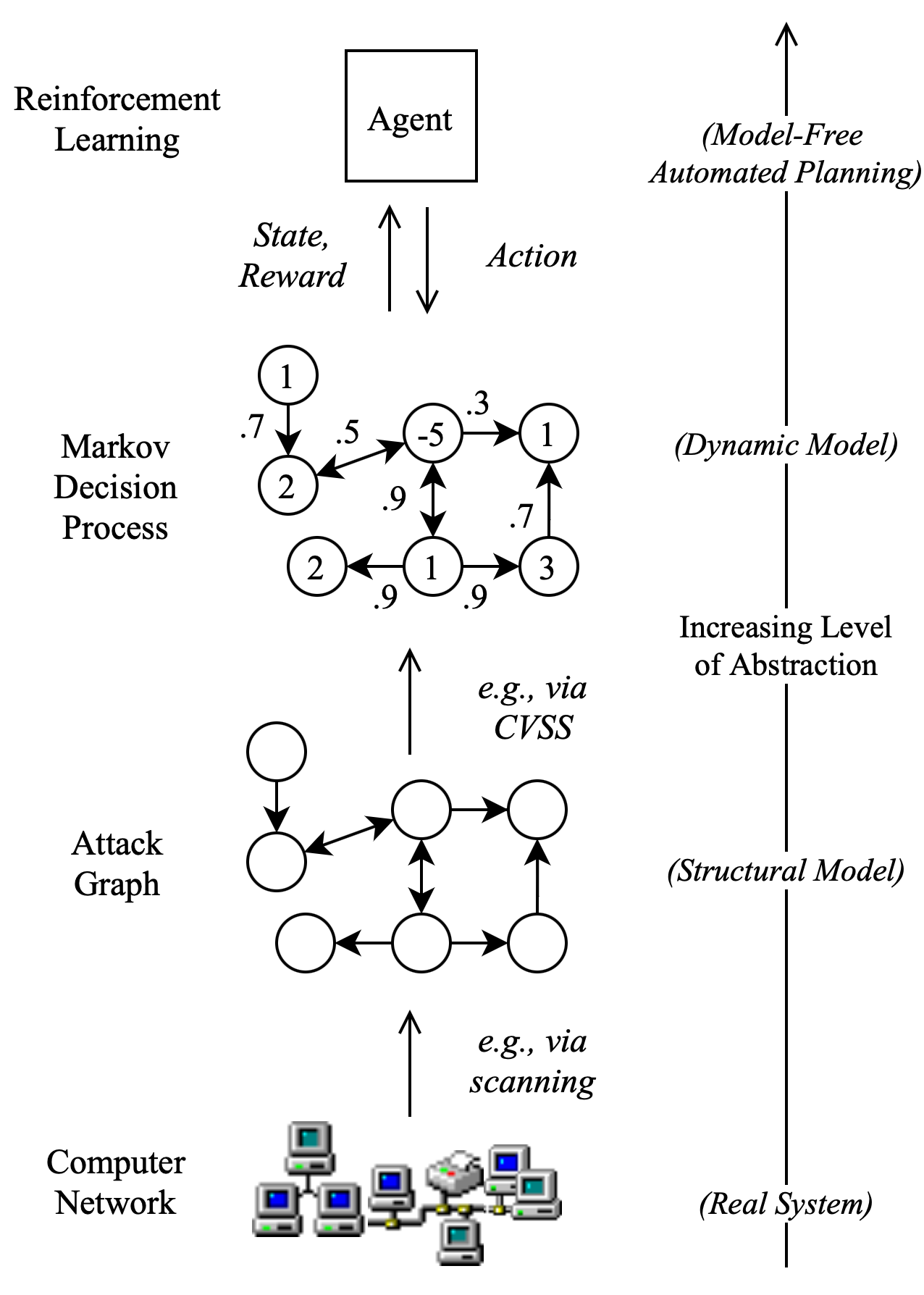}
    \caption{RL for penetration testing requires abstracting from real computer networks described by information such as packet flows, into the mathematical models with which RL agents interact.}
    \label{fig:process}
\end{figure}

While deep RL has been applied to cybersecurity broadly \cite{nguyen2019deep}, it has only recently been pursued as a tool for penetration testing \cite{ghanem2018reinforcement, schwartz2019autonomous, ghanem2020reinforcement, chaudhary2020automated, yousefi2018reinforcement, chowdhary2020autonomous, hu2020automated, gangupantulu2021using, gangupantulu2021crown}. While approaches and uses vary greatly, many use \emph{attack graphs} to model the network \cite{mcdermott2001attack}. Note, attack graphs model the network formed by computer vulnerabilities and exploits, creating an abstraction that does not necessarily match the topology of the physical network, as shown in Figure \ref{fig:process}. The use of attack graphs is a distinguishing character of RL for penetration testing from RL for cybersecurity broadly \cite{nguyen2019deep}.

Most frequently, RL is tasked with simply traversing a network from one initial node to one terminal node---(i.e., finding paths through networks) \cite{ghanem2018reinforcement, schwartz2019autonomous, ghanem2020reinforcement, chaudhary2020automated, yousefi2018reinforcement, chowdhary2020autonomous, hu2020automated, gangupantulu2021using}. Gangupantulu et al., in contrast, emphasize a more complex task of using RL to perform CJ analysis \cite{gangupantulu2021crown}. Here, similar to Gangupantulu et al., the presented RL method solves a more complex task and serves as a targeted tool for cyber operators to improve the efficiency of operator workflow in penetration testing. It does not automate exfiltration entirely.


RL for penetration testing has made frequent use of DQN \cite{schwartz2019autonomous, chowdhary2020autonomous, hu2020automated, gangupantulu2021using, gangupantulu2021crown}. Nguyen et al., alternatively, propose an RL-based approach to penetration testing that uses two agents: one for iteratively scanning the network to build a structural model and another for exploiting the constructed model \cite{nguyen2021proposal}. In our first attempt at performing RL on the network presented later on, we attempt to use the DQN solution method but it did not converge, leading us to explore alternative agents. We use Nguyen et al.'s double agent method where both agents are A2C. We compare to the standard A2C algorithm.

\section{MDPs from Attack Graphs}

There are many solution methods for modeling attack graphs \cite{gonda2018analysis}. Key trade-offs relate to scalability, observability, accuracy, and reliability. In particular, partially observable Markov decision processes (POMDPs) are well-argued to be a more realistic representation of computer networks than MDPs \cite{miehling2018pomdp}. In POMDPs, actions are stochastic and network structure and configuration are uncertain. But POMDPs have not been shown to scale to large networks and require modeling many prior probability distributions \cite{shmaryahu2016constructing}. Additionally, while RL for MDPs is well-established, RL for POMDPs is still under fundamental development \cite{zhu2017improving}. Currently, MDPs are the standard in RL for penetration testing \cite{ghanem2018reinforcement, schwartz2019autonomous, ghanem2020reinforcement, chaudhary2020automated, yousefi2018reinforcement, chowdhary2020autonomous, hu2020automated, gangupantulu2021using, gangupantulu2021crown}.

The CVSS \cite{mell2007complete, joh2011defining} is used as a scalable approach for adding behavior to attack graphs \cite{gallon2011using, keramati2013cvss}. It is the numerical representation of an information security vulnerability. These scores represent an attempt at providing a standardized way of evaluating the severity of threats posed by a particular vulnerability. This takes into consideration both how easy it is to exploit this vulnerability, and also how severe the consequences of such an exploit would be.

While some authors in RL for penetration testing use alternative methods \cite{ghanem2018reinforcement, schwartz2019autonomous, ghanem2020reinforcement, chaudhary2020automated}, CVSS is emerging as a standard approach to modeling MDPs over attack graphs for RL \cite{yousefi2018reinforcement, hu2020automated, chowdhary2020autonomous, gangupantulu2021using, gangupantulu2021crown}. Gangupantulu et al. draw from the literature to define a \emph{vanilla} CVSS-MDP for point-to-point network traversal \cite{gangupantulu2021using}.

CVSS-MDPs use the attack graph to define the state-action space $S \times A$ and CVSS to define the reward $R$ and transition probabilities $P$ \cite{gangupantulu2021using}. CVSS-MDP assigns transition probabilities $P$ using the attack complexity, where the ranks low, medium, and high are associated with transition probabilities of $0.9$, $0.6$, and $0.3$. The reward for arriving at a host is given by,
$$\emph{Base Score} + \frac{\emph{Exploitability Score}}{10}.$$
The agent receives $-1$ reward for each step and receives $100$ reward for arriving at the terminal node. Episodes terminate when the terminal state is reached after number of steps (i.e., actions).

While CVSS scores are useful in practice and currently considered an industry standard, it is important to remember that a measure of threat severity is not the same as a measure of risk and that they do not generalize to give information that's useful for evaluating an entire attack path through a network. From the perspective of an attacker, a greater risk means a greater chance of detection. While the CVSS scores of vulnerabilities do inform the probability of success of any particular exploit in the models here, the real driving force of RL agent behavior should be centered around concepts of terrain \cite{conti_raymond_2018}. The details of this reward engineering of terrain are given in the following.

\section{Methods}

The following subsections describe the presented methods for adding service-based risk penalties as defensive terrain in CVSS-MDPs and the algorithm for discovering the top-$N$ exfiltration paths in a network, shown in Algorithm 1.

\subsection{Defensive Terrain in CVSS-MDPs}

Gangupantulu \textit{et al.} argue that models of cyber terrain can be layered onto CVSS-MDPs, and do so by adding firewalls between subnets, and assigning protocol-specific negative reward and transition probabilities for traversing firewalls \cite{gangupantulu2021using}. Gangupantulu \textit{et al.} later layer on additional notions of cyber terrain by using RL as part of a methodology for modeling footholds and pivot points nearby the 2-hop network of CJ nodes \cite{gangupantulu2021crown}.

We propose a new approach for modeling service-based defensive terrain in CVSS-MDPs. Instead of explicitly defining the defenses in the states of the MDP, we make assumptions similar to what a human attacker would make: even if the attacker cannot detect a defense directly, by their experience they can infer the presence of defenses based on the services available on a given host. Common network defenses can include host-based antivirus and malware detection software, inter-subnet router firewalls, or authentication log tracking.

We engineer rewards for defensive terrain that are additive, or, otherwise put, are layered on top of the CVSS-MDP rewards. A quantified negative reward structure is used to itemize the cost of attacker actions. The criteria of interest are (1) \emph{a risk hierarchy applied to service categories} and (2) \emph{the type of action performed by the agent on a host}. The requirement to implementing these criteria is to unify them in a way the agent could enumerate. This is achieved by creating an array of actions and services and applying an individual reward to each combination. The negative reward can be assigned using (1) action type, here, exploiting or scanning, and (2) services.

Services are derived from four principal categories: authentication, data, security, and common. To create a negative reward, a hierarchy of costs associated with attacking these services was applied. When performing an exploiting action, this hierarchy applies authentication as a reward of -6, data as a reward of -4, while both security and common have a reward of -2. When performing a scanning action, the reward is increased by 1 (i.e., -5, -3, -1, respectively). These rewards represent a combination of factors highlighting the risk to organizations presented from these services. Different organizations or operators may prefer a different scaling. It is important to note that the values of these negative rewards are relative, and as such they can be as a set scaled together to represent different risk preferences. When taking an action on a host with multiple services, the agent applied the highest cost to the action's reward. This was a decision that presumes a leading practice approach by security practitioners to apply security controls based upon the `riskiest' service found on a host (i.e., a service known to be at greater exposure to the network edge or greater business loss if exploited). By syncing our rewards to this presumption, the agent calculates a more realistic quantitative measurement of risk as it attempts to converge to an optimal attack path.

\subsection{Discovering Exfiltration Paths with RL}

\begin{algorithm}[t]
\caption{Exfiltration Paths via RL (EP-RL)}\label{alg:eprl}
\begin{algorithmic}
\Require MDP, initial node $i$, exit nodes $J$, $N$
\Ensure $N$ paths from initial node to top-$N$ exit nodes
\For{$i$ in N}
    \State $path \gets f_{RL}(MDP, i, J)$ \Comment{Optimal path $i \to j, j \in J$}
    \State $paths \gets store(path)$
    \State $J \gets J \setminus j$ \Comment{Remove $j$ from $J$}
\EndFor \\
\Return $paths$
\label{alg:eprl}
\end{algorithmic}
\end{algorithm}

In contrast to Gangupantulu et. al.'s CJ analysis method \cite{gangupantulu2021crown}, our discovering exfiltration paths method uses multiple terminal states corresponding to the various exit nodes of interest and only a single initial node. The agent then interacts with the network in an episodic fashion to learn which is the best exit node with respect to expected reward. To provide a comprehensive path analysis for cyber operators using the tool, the top-$N$ exit nodes are found by iteratively solving the MDP to find the best exit node, removing the best exit node, and solving the MDP again. This algorithm is described in Algorithm 1. Notably, the agent iteratively solves the problem of finding a path to a single exit in the joint set of exits. This avoids the brute force approach of creating an MDP for each exit node, solving each MDP, then ranking the paths.







\section{Experimental Design}

In the following subsections the network, state-action space, and RL algorithm implementation are described.



\subsection{Network Description}

The experimental network where the simulations are run was derived from an architectural leading practices approach to represent enterprise network configurations and deployments. The network contains:
\begin{itemize}
    \item Defined Subnets - 9
    \item Defined Hosts - 26
    \item Types of Operating Systems - 2
    \item Privilege Access Levels - 3
    \item Network Services - 9
    \item Host Processes - 6
    \item Network Firewall Rulesets - 39
\end{itemize}
\noindent The network is visualized in Figures \ref{fig:paths1}, \ref{fig:paths2}, and \ref{fig:paths3}.


Subnets are constructed to represent a grouping of hosts with commonly segregated services utilized for enterprise information technology administration to include server services, database services, client workstation networks, edge and DMZ services, and core services that orchestrate least-privilege or zero-trust security (i.e., domain controllers and public-key infrastructure) \cite{Microsoft:LeastPriv}. Hosts and network firewall rulesets are configured to deliver a representation of common ITS communication requirements between these subnets that allow daily functions of an enterprise ITS department and business operations.

The services within this network are laid out with the presumption of common security controls and monitoring software one would see within an enterprise network. These presumptions include the following expectations:
\begin{enumerate}
    \item Authentication services are exposed to the internet through a Virtual Private Network (VPN).
    \item Web services are exposed to the internet through a secured edge network zone (DMZ).
    \item Services exposed to the internet are monitored.
    \item Firewalls are monitored at a higher rate than other network devices.
    \item Security services have the most inherited security controls.
    \item Authentication services and firewall services, if exploited, have the greatest secondary and tertiary impacts to a network's overall security profile.
    \item Network security rules only apply allowlists.
    \item Host and network assets apply principles of least-privilege when authorizing privileges for account access and use.
\end{enumerate}

\subsection{Environment Description}

For the environment, each host is represented by an 1D vector that contains its status (compromised, reachable, discovered or not) and configurations (address, services, operating system and processes). The environment combines all the vectors for hosts in the network as a entire state tensor. Thus, each state contains descriptions of all hosts. The actions are defined as an operation performed on a specific target host. The actions consist of 6 primitive actions for scanning, exploiting, or privilege escalation. The action type and target host configuration must align or the action will fail. For the environment, the initial host for exfiltration is set on (6, 0), while the terminal hosts are set on (1, 0), (2, 0) and (4, 0), which are all connected to the public internet, where ($a$, $b$) denotes host $b$ in subnet $a$. The initial node is set as compromised, reachable and also discovered at the beginning to make it possible for the agent to perform further actions. The exfiltration goal is reached if the agent compromises any host among them and obtains root access. If the goal is reached, the agent is given a high reward (set as 100 for our experiment).

\subsection{RL Implementation}

The experiment is conducted based on two models: A2C model and the double agent architecture \cite{nguyen2021proposal}. Both agents in double agent use the A2C algorithm. For both, the learning rate is set as 0.001 and the discounted factor is set as 0.99. We use Adam as the optimizer of our networks. Both of the models are trained for 4,000 episodes with a maximum of 3,000 steps in each episode. If the maximum number of steps is reached, the episode terminates and the agent receives 0 terminal reward. Both the A2C model and the structuring agent of double agent use deep neural networks (DNNs) with three fully connected layers of size 64, 32, and 1 and the exploiting agent of the double agent uses a DNN with two fully connected layers of size 10 and 1. All DNNs use tanh activation functions for non-output layers and softmax for the output layer.

\subsection{Sensitivity Analysis}

The experiments run the A2C and double agent algorithms to convergence. To study the effect of the scale of service-based penalties on the convergence of the agents and on the paths they discover, the exploiting and scanning service-based penalties are scaled by a factor of 1.3, 1.0, and 0.7. These values correspond to risk preferences that we term risk-averse, risk-neural, and risk-accepting, respectively.

\section{Results}


To observe the convergence of our models, we plot the steps and the reward versus episodes and the result is shown in Fig. 2. It can be observed that both of our models converge within 1,000 episodes. It could also be noticed that double agent converges slower than the A2C agent, which is expected considering that the double agent is more complex and contains two A2C models that learn simultaneously.

\subsection{RL Performance}

\begin{figure*}[t]
\centering
\includegraphics[width=.36\textwidth]{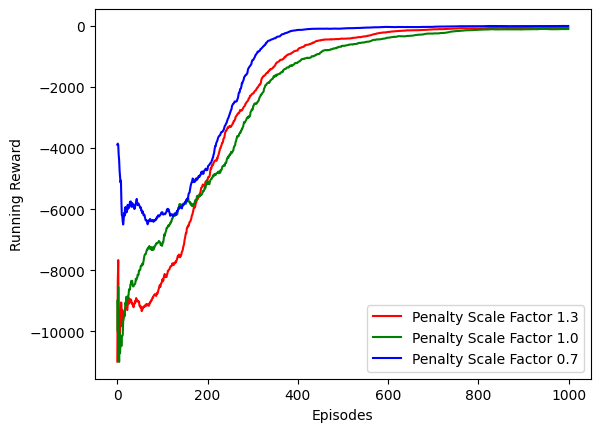}
\includegraphics[width=.35\textwidth]{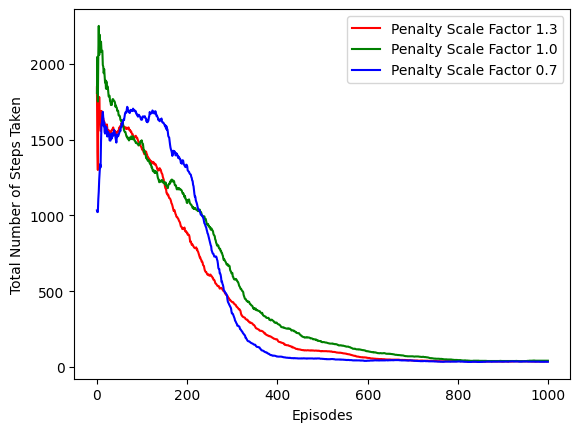}\par
\includegraphics[width=.36\textwidth]{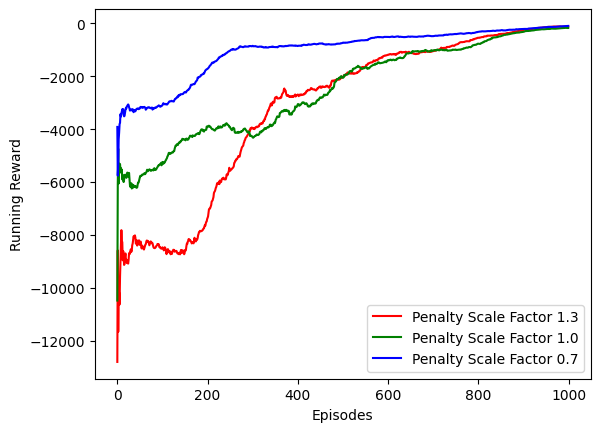}
\includegraphics[width=.35\textwidth]{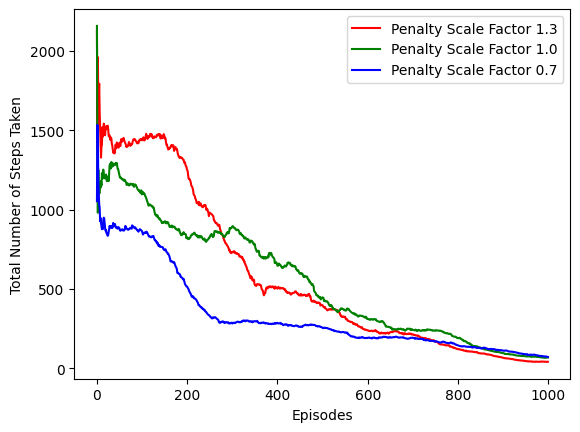}



\caption{Agent learning over episodes. The left plots show the average reward over episodes and the right plots show the average number of steps taken over episodes. The top plots are the results from running RL using an A2C agent and the bottom plots show the results when leveraging the double agent methodology \cite{nguyen2021proposal}. The color of the lines reflects how heavily incentivized an agent is to avoid detection: blue for risk-accepting, green for risk-neutral, and red for risk-averse.}
\label{fig:RL_plots}
\end{figure*}

\begin{table*}[t]
\centering
\begin{tabular}{ cccrrl } 
\hline
Path Rank & Scale Factor & Path & Steps & Reward & Cumulative Probability Score\\
\hline
\multirow{3}{4em}{Best Path} & 0.7 & $(6, 0) \to (3, 0) \to (2, 0)$ & 12 & 57.8 & $2.9+2.9=5.8$ \\ 
& 1.0 & $(6, 0) \to (3, 2) \to (1, 0)$ & 11 & 62 & $2.9+2.9=5.8$ \\ 
& 1.3 & $(6, 0) \to (3, 0) \to (1, 0)$ & 5 & 68.3 & $1.9+2.9=4.8$\\ 
\hline
\multirow{3}{4em}{Second Best Path} & 0.7 & $(6, 0) \to (3, 2) \to (1, 0)$ & 19 & 46.9 & $2.9+4.9=7.8$ \\ 
& 1.0 & $(6, 0) \to (3, 0) \to (2, 0)$ & 16 & 24 & $2.9+4.8=7.7$ \\ 
& 1.3 & $(6, 0) \to (3, 2) \to (1, 0)$ & 19 & 33.1 & $1.9+2.9=4.8$\\ 
\hline
\multirow{3}{4em}{Third Best Path} & 0.7 & $(6, 0) \to (3, 0) \to (1, 1) \to (4, 0)$ & 15 & 41.3 & $1.9+1.9+2.4=6.2$ \\ 
& 1.0 & $(6, 0) \to (3, 2) \to (1, 0) \to (4, 0)$ & 24 & 17 & $3.9+1.9+7.5=13.3$ \\ 
& 1.3 & $(6, 0) \to (3, 2) \to (1, 0) \to (4, 0)$ & 22 & -6.1 & $1.9+2.9+6.3=11.1$\\ 
\hline
\end{tabular}
\caption{Table of the top-3 exfiltration paths found by double agent. \emph{Scale factor} denotes the risk-accepting (0.7), risk-neutral (1.0), and risk-averse (1.3) scaling of the penalty for services. \emph{Path} gives the shortest path from the initial node to exit node, and is derived from the set of actions taken by the converged agent in an episode. \emph{Steps} and \emph{reward}, in contrast, refer to the optimal performance of the agent in an episode (i.e., not just the actions taken to form the \emph{path}). \emph{Cumulative probability score} reports a custom, CVSS-like vulnerability scoring of the \emph{path}.}
\label{table:1}
\end{table*}

When reviewing the A2C and double agent as they reach convergence, A2C uses a similar amount of episodes to reach an optimal path regardless of the scaling factor. In the double agent model, the risk-accepting agent reaches an optimal path much quicker than the risk-neutral and risk-adverse agents. Additionally, the double agent model converges quickly at first, and then plateaus. This suggests the double agent model can quickly arrive at near optimal policies.

\begin{figure}[t]
    \centering
    \includegraphics[width=0.45\textwidth]{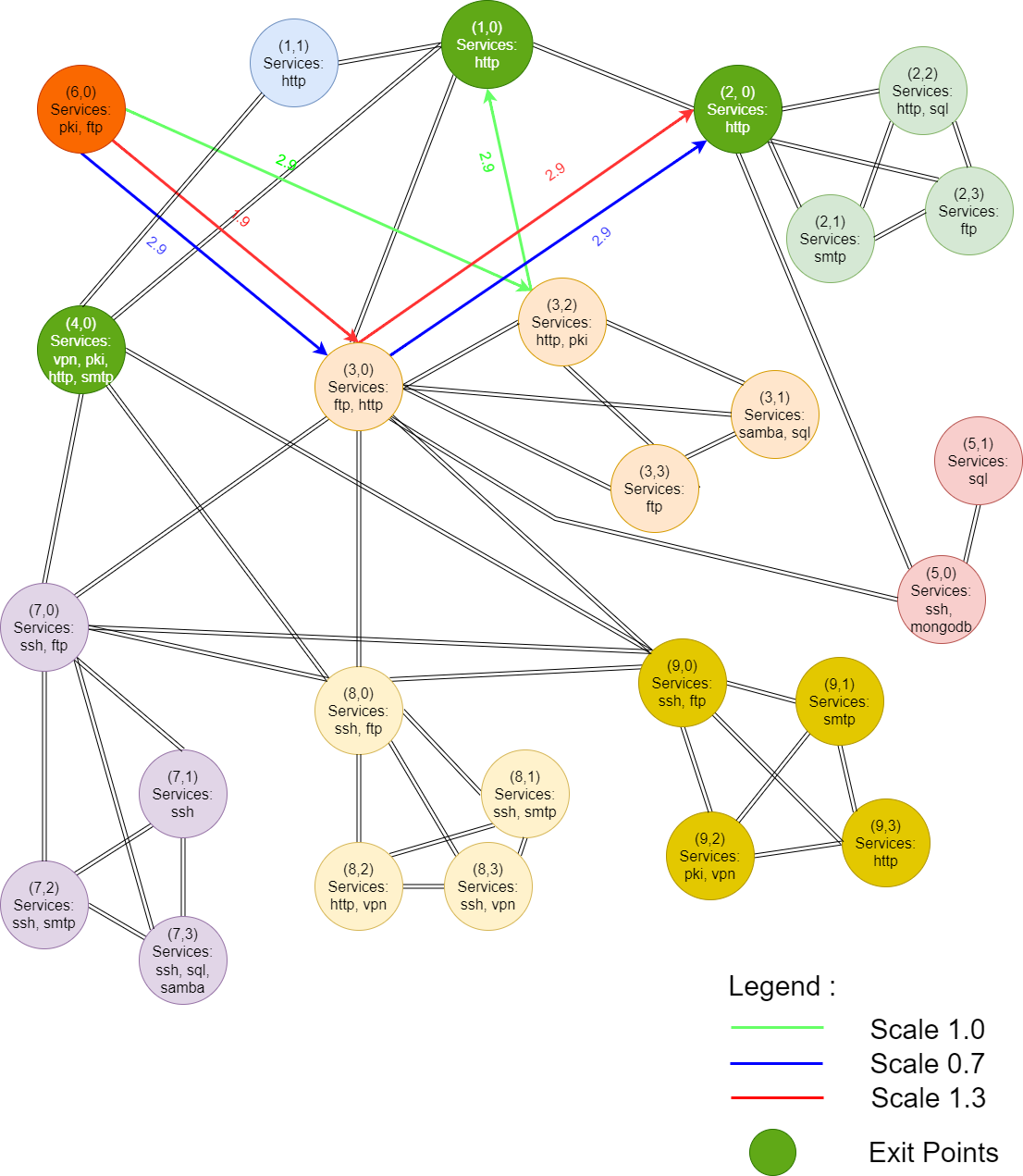}
    \caption{Network diagram showing \emph{Best Path} in Table \ref{table:1}. The color of the edge reflects risk preference and the color of nodes encodes the subnet.}
    \label{fig:paths1}
\end{figure}

\begin{figure}[th]
    \centering
    \includegraphics[width=0.45\textwidth]{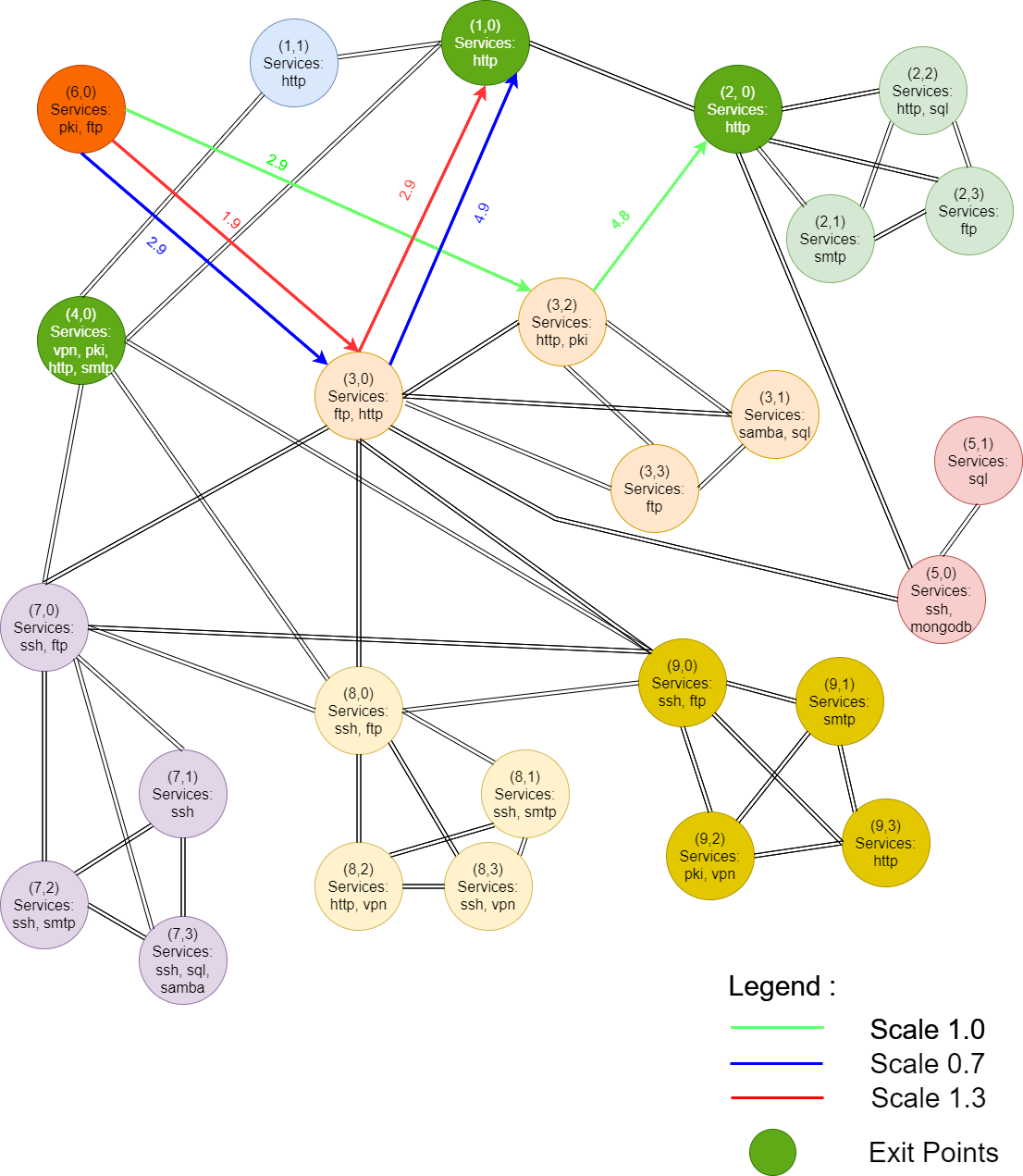}
    \caption{Network diagram showing \emph{Second Best Path} in Table \ref{table:1}. The color of the edge reflects risk preference and the color of nodes encodes the subnet.}
    \label{fig:paths2}
\end{figure}

\begin{figure}[th]
    \centering
    \includegraphics[width=0.45\textwidth]{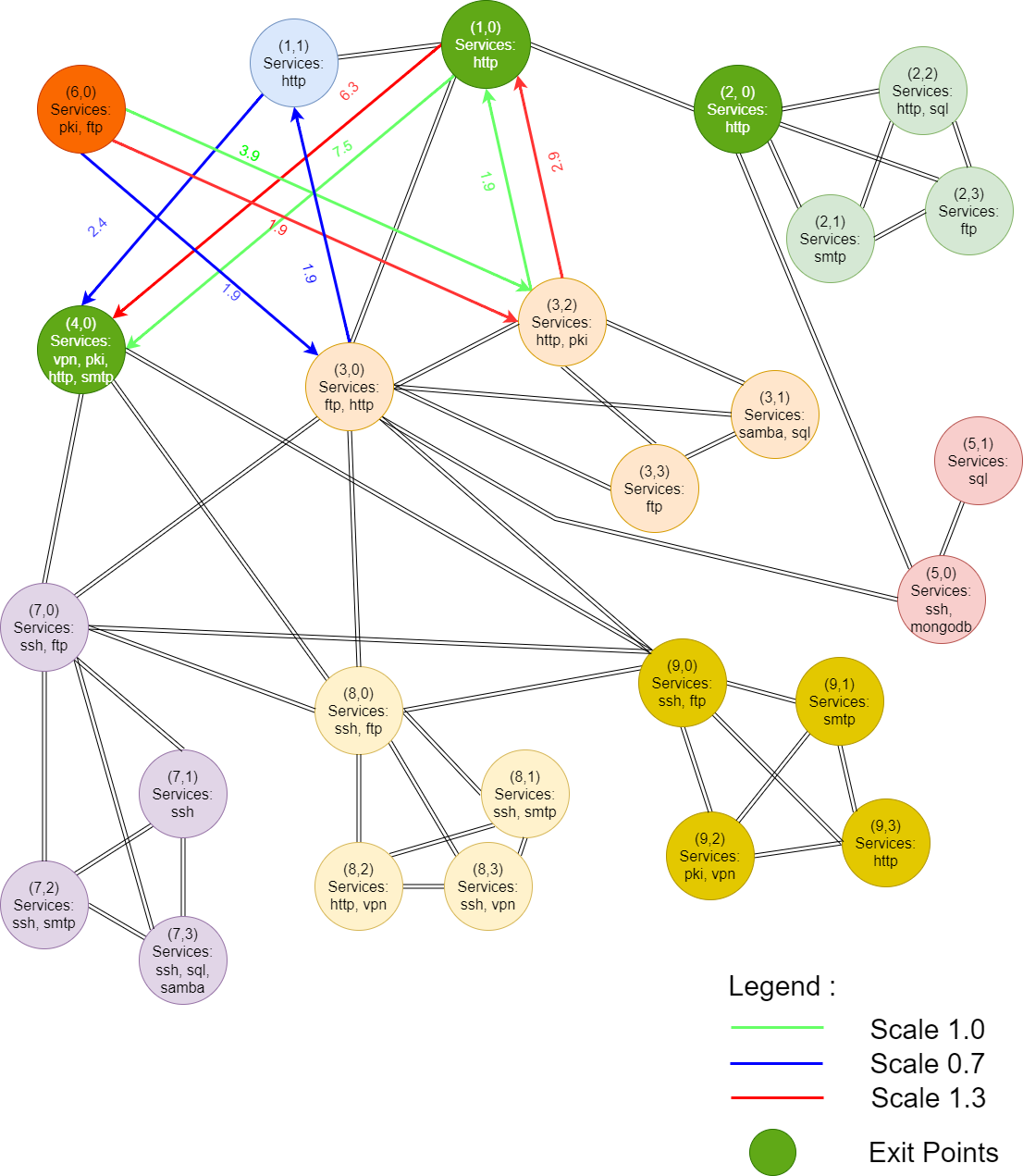}
    \caption{Network diagram showing \emph{Third Best Path} in Table \ref{table:1}. The color of the edge reflects risk preference and the color of nodes encodes the subnet.}
    \label{fig:paths3}
\end{figure}

\subsection{Agent's Behavior Compared to Human Expectations}

The paths results from the nine experiments are shown in Table I. The simulations represent the top three paths for all three risk preferences. In addition to the paths, the table shows the number of steps, the reward and the cumulative probability score. These cumulative probability score are not directly the CVSS scores of the exploits, but are proprietary scores designed to play similar role. It is important to note that the ordering of these scores is \emph{not} the same as the ordering of the reward. This is in agreement with the expectation that the measure of risk (tracked by the reward) does not have to track linearly with a vulnerability score.

Unbeknownst to the rest of the authors, the cyber security operations expert who crafted the simulated networks for the generated attack paths and network topology included two intentional misconfigurations within the host-service assignments. These misconfigurations simulated real-world experiences resolving enterprise network incidents where exfiltration of data occurred. The primary goal of these misconfigurations is to represent flaws in network design that were exploited by actual attackers for exfiltration in actual enterprise networks. If the agents can deduce (without explicit design) this misconfiguration, it would be a compelling example of how this reward engineering can produce human like behavior.

The significant configuration within our results was on host (3,2). The PKI service was extended into subnet (3), the server subnet. In published leading practices for securing PKI \cite{Microsoft:LeastPriv}, this service is included in the most privileged tier of an environment and requires a specific privileged account authorization 'Crytpographic Operator' \cite{Microsoft:LeastPriv}. As such it should only be accessible utilizing hardware and software with enhanced security controls. These leading practices also require this service to only reside in a secure subnet alongside other servers and appliances with similarly privileged security requirements. When this service is allowed to operate as a node within the general server subnet (i.e., regular business applications), it exposes the network firewall rules to exploitation when exfiltrating data from the private key repository database.  


In the resulting optimal path diagrams identified by the agent, host (3,2) was the most traversed node. Reviewing this result shows that the misconfiguration was successfully identified and exploited by the agent when defining an optimal path.

\section{Discussion}

\subsection{Benefits of Approach}

The presented methods can provide security defenders and operators three immediate benefits:
\begin{enumerate}
    \item Iterate security control implementations within enterprise networks by \emph{prioritizing the most impactful controls first}.
    \item \emph{Quantify decreased risk factor} for each iteration of new security controls via the reward.
    \item Deliver integrity to the results by \emph{matching the attacker actions taken to expected actions for each risk preference}.
\end{enumerate}

This RL approach associates to the integrity component of the cybersecurity CIA triad (confidentiality, integrity, and availability). Upon completion of the modeling simulations, the results were analyzed by a cybersecurity operations expert with certification in security architecture and experience resolving incident response from nation-state and APT attack groups. This review found that experiment results matched the expected results for the simulated networks. Relevant criteria for this decision include:
\begin{itemize}
    \item Risk-adverse agent takes very few steps when the entire network is exposed.
    \item Risk-accepting agent will achieve a greater reward in more secure networks because of its ability to move faster than stealthier actors.
    \item The optimal path will often be the same for various risk profiles, matching the A2C modeling convergence trends.
    \item Utilizing misconfigurations of security services within a network is a high-likelihood of success for attackers.
    \item Data exploitation is more likely through servers and services than through client workstations.
    \item When operating in more secure networks, the agent consistently creates simple exfiltration paths but requires additional unsuccessful scanning actions to achieve this same convergence.
\end{itemize}

\subsection{Remarks on Payload}

Within the current scope of this work, there is no consideration for the size of the payload extracted, or the rate at which it is removed. If the payload is a small amount of (critical) data, this simulation can be considered an approximation of reality. If the amount of data exfiltrated becomes large enough that this approximation fails, then additional modeling considerations need to be considered, such as encoding rates of transfer and amount of data into the states of the environment. Payload size, while being a calculable statistic for security operations, is often measured for security in a binary manner. If the payload size from one server to another server, or for one firewall at a given time of day, is of sufficient variation from the expected thresholds, an alert will trigger for security or network operations. While malicious actions can create this, non-malicious actions can create this as well. Common ITS operations such as database backups, system update downloads, or unexpected network configuration changes can each create a pattern that alerts security or network teams to heavy payloads on a network. Without a way to compensate for these additional variables, the value of adding payload sizes in this work was negligible.

\section{Conclusion}

In this paper, we have provided security practitioners and network defenders a quantitative methodology using RL to identify optimal paths for  data exfiltration. In our experiments, the presented RL approach identified the most likely hosts and services used when exfiltrating data and captured metrics used in network risk assessments. The strength of this approach was validated through identification of intentional network misconfigurations that mimic real-world vulnerabilities. 



Future work should consider integration with other RL for penetration testing tasks. In addition, expanding the risk formalism to increase its sophistication and maturity will drive increased applicability and relevance. Review of payload size extraction and subsequent rates are also should also be included for future studies.

\section{Acknowledgements}

This work was made possible through the collaboration between Dr. Michael Ambroso, Lead for the AI/ML for Cyber Testing Innovation Pipeline group---funded by Deloitte's Cyber Strategic Growth Offering led by Deborah Golden, and Dr. Laura Freeman, Director of the Hume Center for National Security and Technology's Intelligent Systems Lab at the Virginia Polytechnic Institute and State University.

\bibliographystyle{IEEEtran}
\bibliography{ref}

\end{document}